\documentclass[12pt,preprint]{aastex}
\usepackage[pdftex]{epsfig}

\newcommand{\Grad}{$^{\circ}$}

\shortauthors{T.~Antoni et al. (KASCADE Collaboration) }
\shorttitle{Search for cosmic ray point sources with KASCADE}

\begin{document}

\title{Search for cosmic-ray point sources with KASCADE}

\author{T.Antoni\altaffilmark{1},
        W.\,D.~Apel\altaffilmark{2},
        A.F.~Badea\altaffilmark{2,6},
        K.~Bekk\altaffilmark{2},
        A.~Bercuci\altaffilmark{2,6},
        H.~Bl\"umer\altaffilmark{2,1},
        H.~Bozdog\altaffilmark{2},
        I.\,M.~Brancus\altaffilmark{3},
        C.~B\"uttner\altaffilmark{1},
        K.~Daumiller\altaffilmark{1},
        P.~Doll\altaffilmark{2},
        R.~Engel\altaffilmark{2},
        J.~Engler\altaffilmark{2},
        F.~Fe{\ss}ler\altaffilmark{2},
        H.\,J.~Gils\altaffilmark{2},
        R.~Glasstetter\altaffilmark{1,7},
        A.~Haungs\altaffilmark{2},
        D.~Heck\altaffilmark{2},
        J.\,R.~H\"orandel\altaffilmark{1},
        K.-H.~Kampert\altaffilmark{1,2,7},
        H.\,O.~Klages\altaffilmark{2},
        G.\,Maier\altaffilmark{2,9},
        H.\,J.~Mathes\altaffilmark{2},
        H.\,J.~Mayer\altaffilmark{2},
        J.~Milke\altaffilmark{2},
        M.~M\"uller\altaffilmark{2},
        R.~Obenland\altaffilmark{2},
        J.~Oehlschl\"ager\altaffilmark{2},
        S.~Ostapchenko\altaffilmark{1,8},
        M.~Petcu\altaffilmark{3},
        H.~Rebel\altaffilmark{2},
        A.~Risse\altaffilmark{5},
        M.~Risse\altaffilmark{2},
        M.~Roth\altaffilmark{1},
        G.~Schatz\altaffilmark{2},
        H.~Schieler\altaffilmark{2},
        J.~Scholz\altaffilmark{2},
        T.~Thouw\altaffilmark{2},
        H.~Ulrich\altaffilmark{2},
        J.~van Buren\altaffilmark{2},
        A.~Vardanyan\altaffilmark{4},
        A.~Weindl\altaffilmark{2},
        J.~Wochele\altaffilmark{2},
        and J.~Zabierowski\altaffilmark{5} \\ (The KASCADE Collaboration)
       }

\altaffiltext{1}{Institut f\"ur Experimentelle Kernphysik, Universit\"at
        Karlsruhe, 76021~Karlsruhe, Germany}
\altaffiltext{2}{Institut f\"ur Kernphysik, Forschungszentrum Karlsruhe,
             76021~Karlsruhe, Germany}
\altaffiltext{3}{National Institute of Physics and Nuclear Engineering,
             7690~Bucharest, Romania}
\altaffiltext{4}{Cosmic Ray Division, Yerevan Physics Institute,
             Yerevan~36, Armenia}
\altaffiltext{5}{Soltan Institute for Nuclear Studies,
             90950~Lodz, Poland}
\altaffiltext{6}{on leave of absence from NIPNE, Bucharest}
\altaffiltext{7}{now at: Universit\"at Wuppertal, 42119 Wuppertal, Germany}
\altaffiltext{8}{on leave of absence from Moscow State University, Moscow, Russia}
\altaffiltext{9}{corresponding author, email: gernot.maier@ik.fzk.de}


\begin{abstract}
A survey of the northern hemisphere for astrophysical point sources with
continuous emission of high-energy cosmic rays is presented.
Around 47 Mio extensive air showers with primary energies above
$\approx$ 300~TeV measured by the KASCADE
detector field are selected for this analysis.
Besides the sky survey, a search for signal excess in the region
of the galactic plane and of selected point source
candidates has been performed.
There is no evidence for any significant excess.
This is valid for an analysis of all recorded showers as well
as for a data set enhanced in gamma-ray induced showers.
An upper flux limit of around $3\times 10^{-10}$ m$^{-2}$s$^{-1}$
for a steady point source that transits the zenith is obtained.
Additionally, the distribution of the arrival directions of extensive air showers 
with energies above 80~PeV were studied by an autocorrelation
analysis.

\end{abstract}

\keywords{cosmic rays; anisotropy; point sources; sky survey; gamma rays}

\section{Introduction}

The origin of high-energy cosmic rays is still unknown.
Their sources are obscured due
to the deflection of the charged particles in 
galactic magnetic fields.
Only neutral particles, i.e.~neutrons or 
high-energy photons, can point back to their sources.
But besides restrictions from the known acceleration processes,
several limitations
for the detection of neutral particles arise from
propagation effects and from the large
background of charged cosmic rays.

Neutrons can reach the Earth if their
energy and hence their decay length is
comparable with the distance of the source.
A decay length of 1~kpc corresponds to a  
neutron energy of about $10^{17}$~eV,
the distance to the galactic center (8.5~kpc)
to an energy of roughly $10^{18}$~eV.
There are models proposing neutron
emission from the center of the galaxy
\citep{Hayashida:1999}.
They could explain the excess of events
from this direction
(which is unfortunately not visible from the \mbox{KASCADE} site)
measured by the AGASA experiment \citep{Hayashida:1999}
at around $10^{18}$~eV.
In the present work the possibility of neutrons
originating from more nearby sources ($\approx 1$~kpc)
is analysed with the data set of KASCADE.

TeV $\gamma$-ray sources are potential sites
for the acceleration of high-energy cosmic rays.
Several sources have been established in the last
decade.
The energy spectrum of some of them extend 
up to 80~TeV (e.g. \cite{Tanimori:1998b}).
Limitations for the detection
of PeV $\gamma$-ray sources are:
very low fluxes due to the steeply falling energy spectra
of the sources,
a possible cutoff in their spectra,
and the attenuation of $\gamma$-rays on their way
to the observer as a result
of the interaction with low energy photons from the 
cosmic microwave background radiation \citep{Coppi:1997}.
A review of the status of high-energy $\gamma$-ray astronomy can
be found in \citep{Ong:1998,Hoffmann:1999}.

All TeV $\gamma$-ray sources have been observed by
imaging atmospheric Cherenkov telescopes. 
While these instruments are very sensitive to
$\gamma$-ray showers (low energy threshold, good rejection
of hadron-induced extensive air showers), the major drawback for
the search of new sources is their small field of
view and their limitation to dark observation periods.
Large field arrays measuring the secondary particle distribution
of extensive air showers on ground have large fields of
view and duty cycles independent of day and night
periods.
Unfortunately, they suffer from
the huge background
of hadron-induced extensive air showers.
Only the Tibet air shower array and the Milagro experiment
reported the detection
of a TeV $\gamma$-ray source (Crab-Nebula)
\citep{Amenomori:1999,Atkins:2003}.

Several sky surveys for steady $\gamma$-ray sources
have been performed by air shower experiments
at energies from 1~TeV to 1~PeV
(CYGNUS \citep{Alexandreas:1991b},
CASA \citep{McKay:1993}, Milagrito \citep{Wang:2001},
Tibet \citep{Amenomori:2002}, 
HEGRA \citep{Aharonian:2002}).
None of the experiments have found point sources.
In this article a survey of the northern sky for primary
energies above 300~TeV in 
the declination range 15\Grad \ $ < \delta <$ 80\Grad \ 
with data recorded by the KASCADE detector field is 
presented.
Compared to the experiments mentioned above,
KASCADE measures in a different energy region or
observes a different part of the sky.
An analysis of the arrival-direction distribution
of charged cosmic rays on large angular scales
covering the effects of their diffusion in the
galactic magnetic field is presented elsewhere
\citep{Antoni:2003c}.


\section{KASCADE - experimental setup and shower reconstruction}

KASCADE (KA{\em rlsruhe} S{\em hower} C{\em ore and} A{\em rray}
DE{\em tector}) is located at Forsch\-ungs\-zentrum Karls\-ruhe,
Germany ($8.4^{\mathrm{o}}$~E, $49.1^{\mathrm{o}}$~N) 
at \mbox{110~m} a.s.l.
corresponding to an average vertical atmospheric depth of \mbox{1022~g/cm$^2$}.
The experiment measures the electromagnetic, muonic, and hadronic
components of extensive air showers with three major detector 
systems: a large field
array, a muon tracking detector, and a central detector.
A detailed description of the KASCADE experiment can be found
in \citep{KASCADE03}.

In the present analysis data from the 200$\times$200~m$^2$ scintillation
detector array are used.
The 252 detector stations are uniformly spaced on a square grid of 13~m.
The stations are organized in 16 electronically
independent so-called clusters with 16 stations in the 12 
outer and 15 stations in the four inner clusters.
The stations in the inner/outer clusters contain four/two
liquid scintillator detectors covering a total area of 490~m$^2$.
Additionally, plastic scintillators are mounted below an absorber
of 10~cm of lead and 4~cm of iron in the 192 stations of the outer clusters
(622~m$^2$ total area).
The absorber corresponds to 20 electromagnetic radiations lengths 
entailing a threshold for vertical muons of 230~MeV.
This configuration allows the measurement of the electromagnetic
and muonic components of extensive air showers.
The detector array reaches full efficiency on the detection of 
showers for electron numbers $\log_{10} N_e > 4$.
The trigger rate is about 3~Hz.

Applying an iterative shower-reconstruction procedure the
number of electrons and muons in a shower are determined
basically by maximizing a likelihood function describing
the measurements with the
Nishimura-Kamata-Greisen (NKG) formula~\citep{Kamata:1958,Greisen:1965}
assuming a Moli\`ere radius of 89~m. 
Detector signals are corrected for contributions of other particles,
i.e. the electromagnetic detectors for contributions of muons,
gammas, and hadrons \citep{Antoni:2001}.
Shower directions and hence the directions of the incoming primary particles
are determined without assuming a fixed
geometrical shape of the shower front by evaluating the arrival times of
the first particle in each detector and the total particle number
per station.

\begin{figure}
\plotone{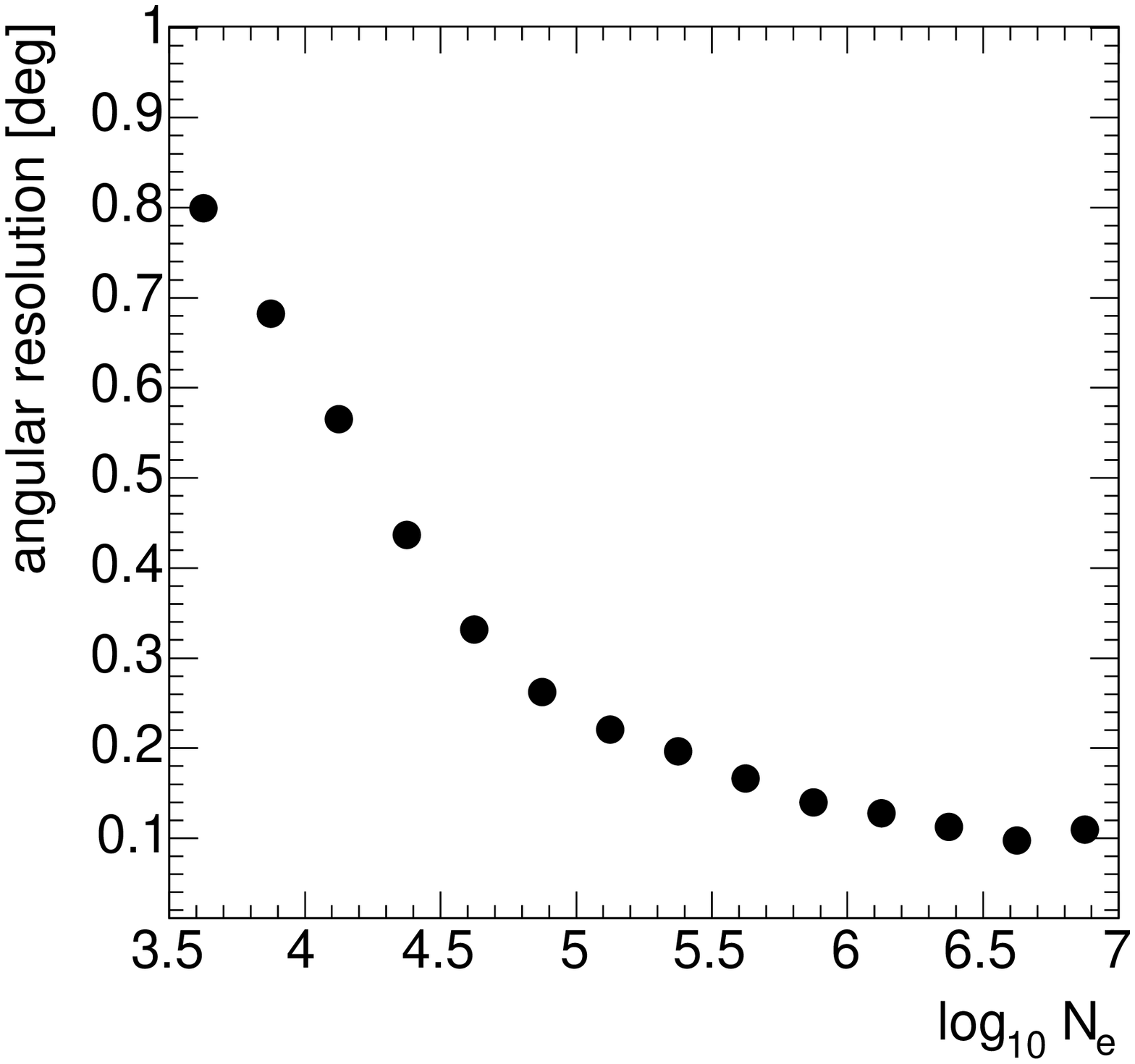}
\caption{\label{fig:reso}
Angular resolution (68\%) of the KASCADE detector field.
Statistical uncertainties are smaller than the marker sizes.}
\end{figure}

The angular resolution of the KASCADE detector field is
determined by the application of the chequer board method.
Dividing the detector field in two subarrays and comparing
shower directions using the one or the other subarray gives
a measure for the angular resolution.
Shower directions are reconstructed with a resolution (68\% value) of
0.55\Grad \ for small showers with $\log_{10} N_e\approx 4$
 and 0.1\Grad \ for showers with
an electron number of $\log_{10} N_e \geq 6$ 
(see Figure \ref{fig:reso}).
The angular resolution is almost independent of the zenith angle at all
shower sizes, e.g. small showers with zenith angles larger
than 35\Grad \ are reconstructed with 
an accuracy of about 0.58\Grad .

Systematic uncertainties in the angular reconstruction have been studied by 
shower simulations and by real shower measurements.
The simulation chain consists of CORSIKA (version 6.014) \citep{Heck:1998}
using the hadronic
interaction models QGSJet \citep{Kalmykov:1997} and GHEISHA
\citep{Fesefeldt:1985}, and 
for the electromagnetic part EGS4 \citep{Nelson:1985}
followed by a detailed simulation of the detector response
based on GEANT \citep{geant}.
After completion of the simulations a GHEISHA version \citep{Heck:2003}
corrected for program mistakes was available,
but recent comparision show no effect of these corrections to
the presented investigations.
No systematic uncertainties are visible neither in simulations
nor in real shower measurements.
In the second method the reconstructed shower direction of the
detector field is compared with the shower direction of the same shower
determined by the muon tracking detector of KASCADE \citep{Doll:2002}.
This component of KASCADE
measures tracks of individual muons ($E_{\mu}>800$~MeV) with an
arrangement of three layers of limited streamer
tubes of 128~m$^{2}$ area each. 
The comparison results in an angular difference
between both reconstructed shower directions
of $\Delta = 0.01$\Grad $\pm 0.03$\Grad .

\begin{figure}
\plotone{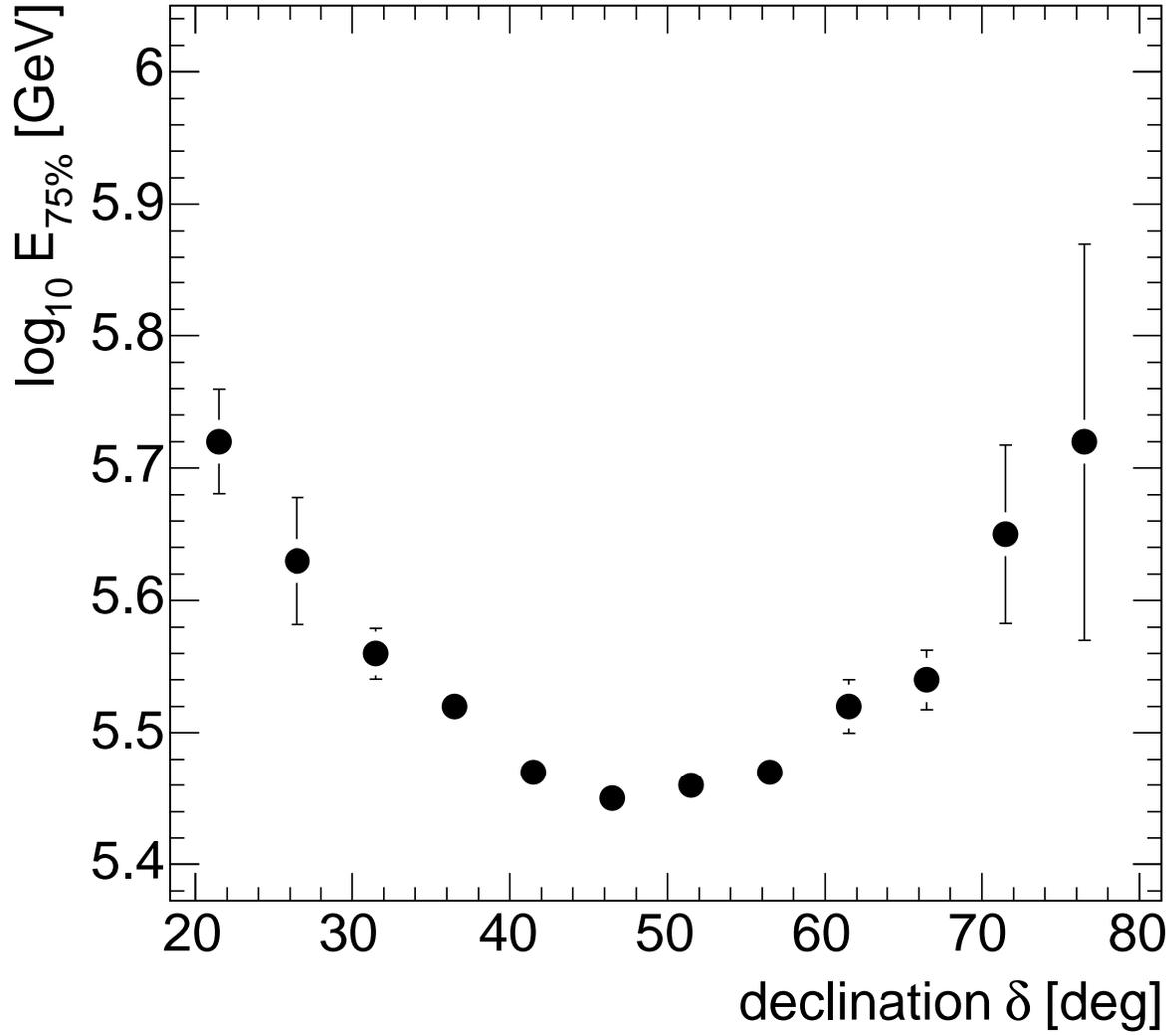}
\caption{\label{fig:thresh}
Energy threshold (75\% detection probability)
of KASCADE
for $\gamma$-induced showers.}
\end{figure}

The energy threshold of KASCADE is defined by the trigger
conditions and the data cuts given below.
It is a function of zenith angle and therefore declination dependent.
Figure \ref{fig:thresh} shows the mean values for 75\% detection probability
of $\gamma$-induced showers vs.~declination.
The realistic zenith angle distributions as function of declination
are taken into account for these calculations. 
The threshold is determined with the already described extensive air shower simulations.
The threshold is increasing from 280~TeV at $\delta=49$\Grad \ to
550~TeV at the edges of the field of view.

\section{Data selection and suppression of hadron induced showers}
\label{sec:gammas}

To clean the data set from poorly reconstructed showers, the following
cuts were applied:
more than 13 out of 16 clusters are working,
shower core positions are inside a circular area of
91~m radius around the center of the array
to omit large reconstruction errors at the edges
of the detector field, and
zenith angles are requested to be $\Theta<$ 40\Grad .

The data set was recorded between May 1998 and October 2002
corresponding to an effective time of about 1300 days.
About 47 Mio events are left for the analysis after the mentioned cuts.

\begin{figure}
\plottwo{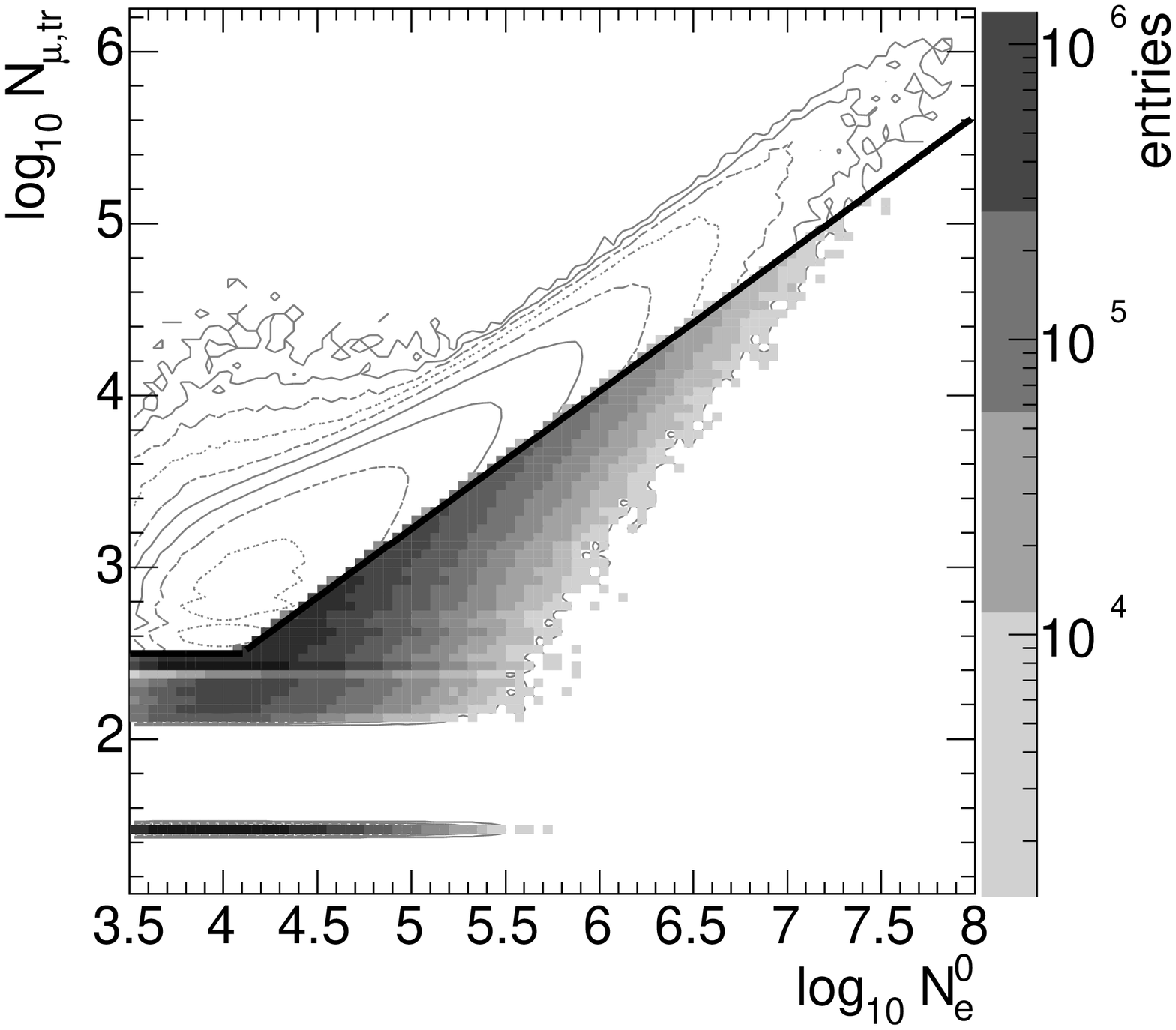}{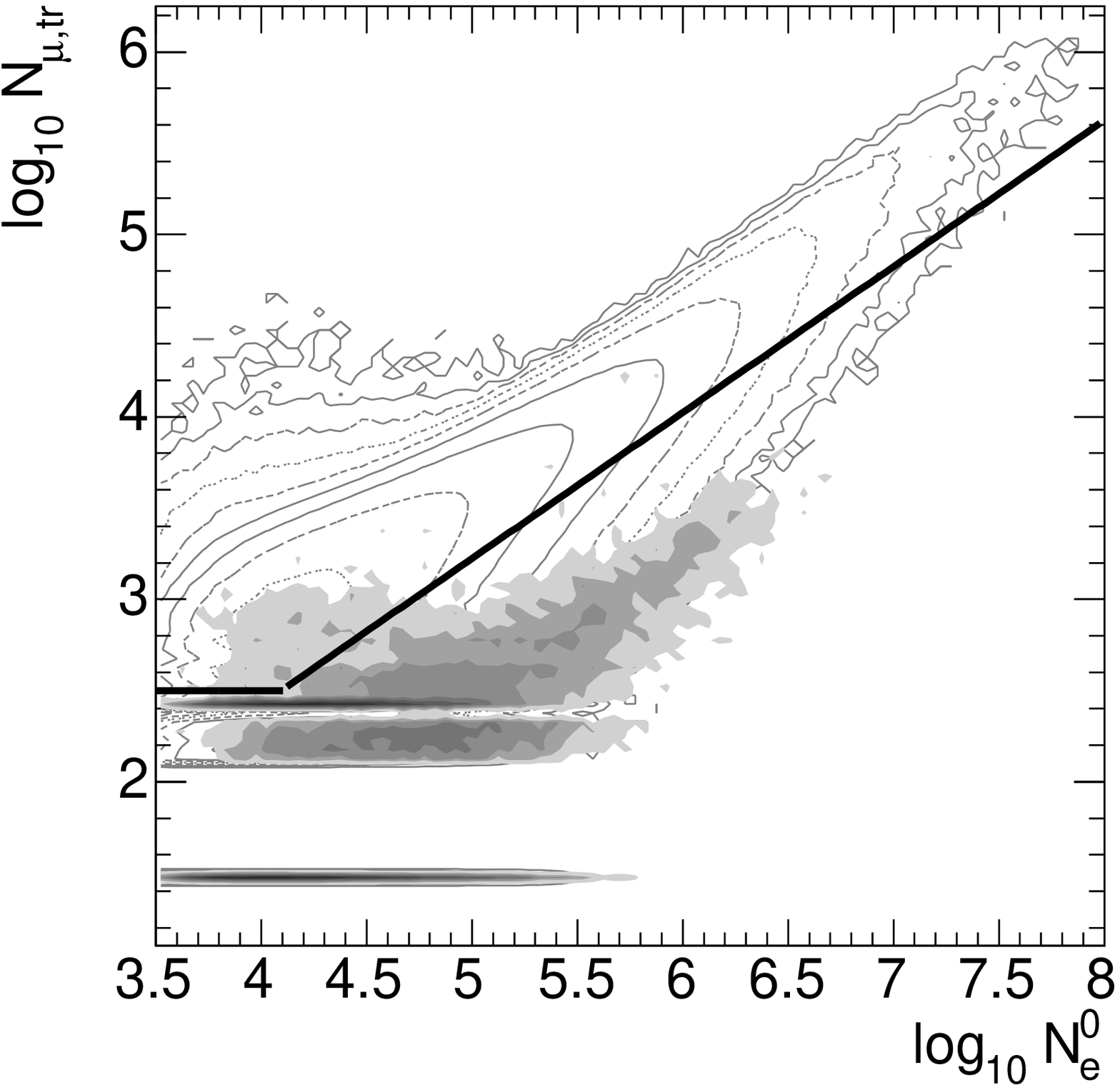}
\caption{\label{fig:gammaSep}
Left-hand side: Number of muons versus number of electrons ($\log_{10} N_{\mu,tr}$ vs. $\log_{10} N_e^0$)
of showers measured with KASCADE.
The contour line histogram indicates the whole data set, the
shaded area illustrates the selection of muon-poor showers.
Right-hand side: Number of muons versus number of electrons for
$\gamma$-induced showers simulated with CORSIKA followed by a detailed
detector simulation.
The contour line histogram indicates again the data set of measured showers.
In both figures the cut to suppress hadron-induced showers is shown
as straight lines.
Showers with no registered muons are plotted as $\log_{10} N_{\mu,tr}=1.5$.
}
\end{figure}

The sensitivity to  $\gamma$-ray induced showers can be enhanced by
the suppression of hadron-induced extensive air showers
using the ratio of number of muons to number of electrons
in a shower. 
Figure \ref{fig:gammaSep} (left-hand side)
shows the distribution of electron vs.~truncated
muon number ($\log_{10} N_{\mu,tr}$-$\log_{10} N_e^0$) for
measured showers.
$N_{\mu,tr}$ denotes the number of muons in the distance range from 
40 to 200~m to the shower core.
The electron number $N_e$ is corrected 
to a zenith angle of $\Theta=0$\Grad \ using an attenuation length
of $\Lambda_{N_e}=175$~g/cm$^2$ \citep{Antoni:2003}.
The $\log_{10} N_{\mu,tr}$-$\log_{10} N_e^0$ distribution
for simulations of $\gamma$-induced showers is shown
in Figure \ref{fig:gammaSep} (right-hand side).
The showers are simulated in the energy range
$5\times 10^{13}$ eV $ < E_0 < 5\times 10^{15}$ eV
following a power law with a spectral index of -2.
The distribution of shower sizes for $\gamma$-induced showers
motivates the following cuts to suppress hadron-induced showers:
$ \log_{10} N_{\mu,tr} < 2.5$ for  $\log_{10} N_e^0 < 4.1$ 
and
$\log_{10} N_{\mu,tr} < -0.78 + 0.8\times \log_{10} N_e^0$
for $\log_{10} N_e^0 \geq 4.1$.
This selection of muon-poor showers is indicated by straight lines in 
Figure \ref{fig:gammaSep}.
About 75\% of the hadron-induced showers are suppressed by this procedure.
All results in the following section are presented for the whole data set
as well as for the data set consisting of muon-poor showers only.


\section{Point source search}

\subsection{General method}
\label{sec:analysis}

A region in the sky of a certain angular size is analyzed by comparing the
number of events from the assumed direction with an expected number of
background events.
Sky regions with significant excesses indicate possible point-like sources.
The significance for the deviations from the expected background is calculated by
the widely used method of \cite{Li:1983}.

The so-called time-shuffling method \citep{Cassiday:1990,Alexandreas:1991} has been used for
the background calculation.
With this method  artificial background events are created with the same
arrival times as the measured events.
The shower directions in horizontal coordinates (azimuth and zenith)
for these new events are taken
from other randomly selected measured events.
In this manner artificial data sets are generated by
many repetitions of this procedure and by usage of the time dependent
conversion from horizontal to equatorial coordinates.
The mean shower-direction distribution from these generated data sets
has most of the properties of an isotropic background.
Exposure, angular distributions and possibly existing systematic reconstruction
errors are the same as in the real data set.
Interruptions in the data acquisition are taken into account as well. 
The loss of sensitivity due to the overestimation of the number of
background events by the usage of possible source events is negligible since
showers are distributed over the whole right-ascension range.
In this analysis the expected background is determined by the average of
50 artificial data sets.

\begin{figure}
\plotone{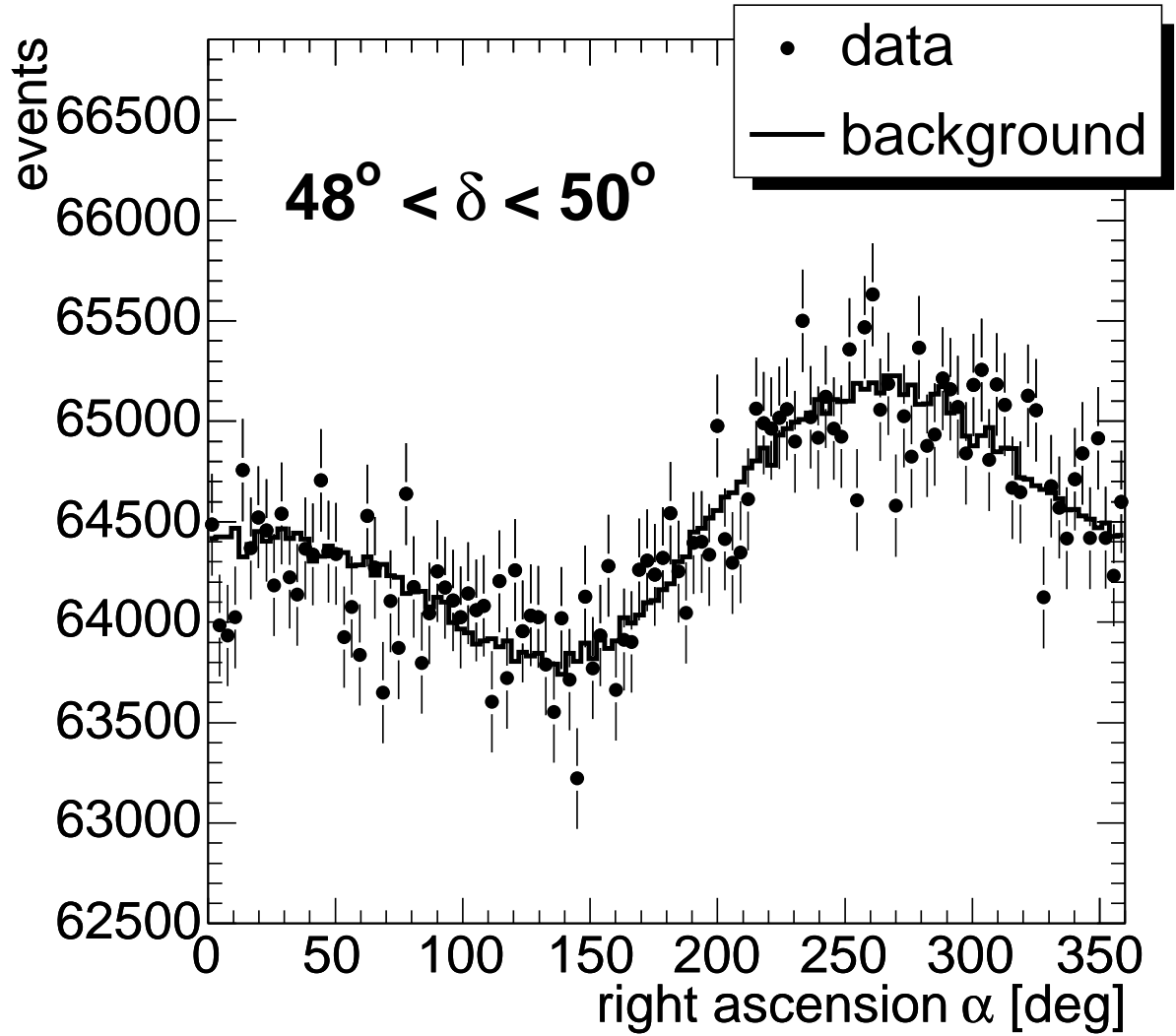}
\caption{\label{fig:UnDetail}
Right ascension distribution for showers in the declination band
48\Grad \ $<\delta <$ 50\Grad.
The background generated by the time-shuffling technique is shown by
the solid line.
The origin of the ordinate is suppressed.
}
\end{figure}

Sky maps in equatorial coordinates of the distribution of arrival directions of
the measured events and the expected background are generated by the time-shuffling technique.
Sky maps of significance values are calculated by comparing the content of each bin
in the data maps with the corresponding value in the background maps.
The bins in the sky maps are of constant solid angle,
the bin width is optimized by Monte Carlo calculations
to give maximum sensitivity to point-like sources.
The solid angle of a bin is selected to be $\Delta \Omega=7.6\times 10^{-5}$~sr with a 
bin width in declination of 0.5\Grad \ and 0.5\Grad $/\cos \delta$ in right ascension.
The bins are independent and non-overlapping.
To avoid the loss of sensitivity to sources located on the edge
of the bins, all analyses are repeated with sky maps of different binning
definitions.
As example, Figure \ref{fig:UnDetail} shows the data and background maps in the declination
range 48\Grad \ $ < \delta < $ 50\Grad.
The smoothness of the generated background distribution compared with the measured data
is clearly visible.
The modulation of the distributions (amplitude $\approx$ 1\%)
is due to interruptions during the data acquisition.

In case of no point-like sources, the distribution of significances in the sky
map is expected to be Gaussian with a mean value $\mu = 0$ and width $\sigma = 1$ \citep{Li:1983}.
Deviations from isotropy would be visible in a non-gaussian shape of the significance
distributions or in the occurrence of values with unlikely large significances.

The 90\% upper limits $N_{lim}$ of events above background
in the source regions are determined by the method of \cite{Helene:1983}.
With the assumption of equal power laws in the energy spectra of background and
source events, the upper flux
limits for showers with primary energies larger than $E_0$ are calculated by:
\begin{equation}
\label{equ:flux}
F_{lim}( >E_0) = \frac{N_{lim}}{N_{B}} \frac{f_{B}(>E_0) \Delta \Omega}{\epsilon}
\end{equation}
$N_{B}$ is the number of background events in the analyzed bin or region, $f_{B}(>E_0)$ the
background flux above a certain primary energy $E_0$, $\Delta \Omega$ the observed solid angle
and $\epsilon$ the average fraction of source events in the search region.
Monte-Carlo simulations of shower directions from a point-like source with 
consideration of the angular resolution of KASCADE yield $\epsilon=(67.6\pm0.5)$\%,
assuming a neglectable error of the source position.

Circular regions around the objects of the current catalog of TeV-$\gamma$-ray
sources are inspected in more detail.
The search radius is 0.5\Grad .
The expected background is generated as described
with the time-shuffling technique using arrival
times and directions of the off-source events.
The significance of the deviations from the expected background as well as upper flux
limits are calculated in the same manner as for the sky survey.

\subsection{Sky survey}
\label{sec:results}

\begin{figure}
\includegraphics[width=0.85\linewidth]{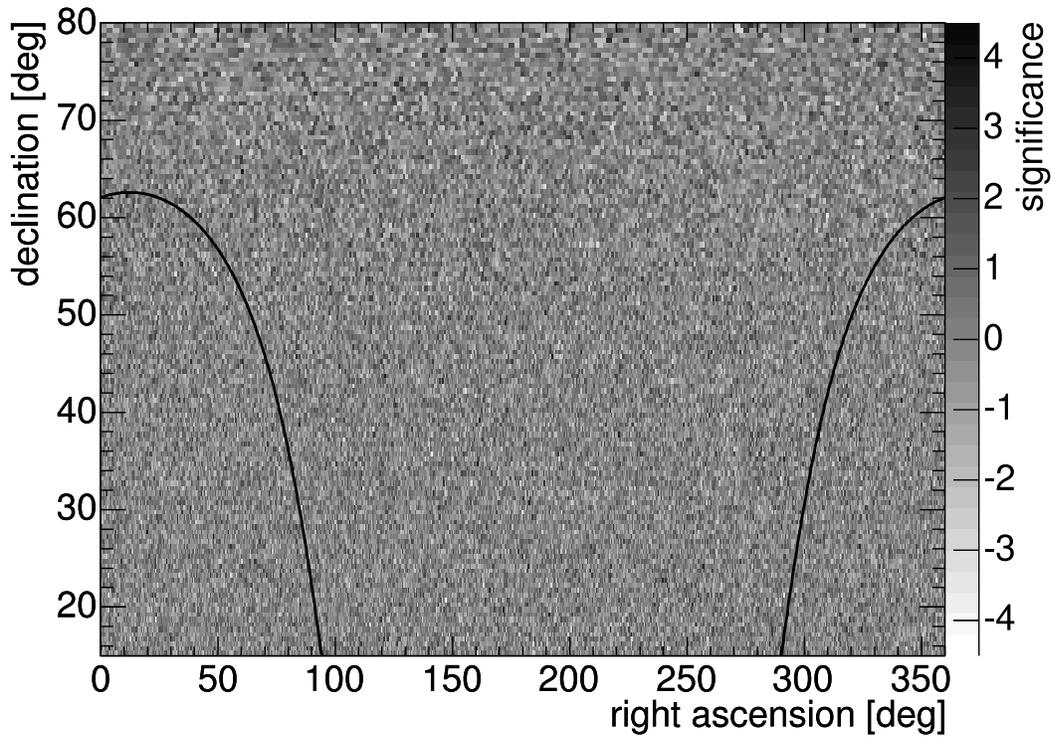}
\caption{\label{fig:LiMa}
Significance distribution in equatorial coordinates
for the data sample of muon-poor extensive air showers.
The galactic plane is indicated by the line.}
\end{figure}

\begin{figure}
\plottwo{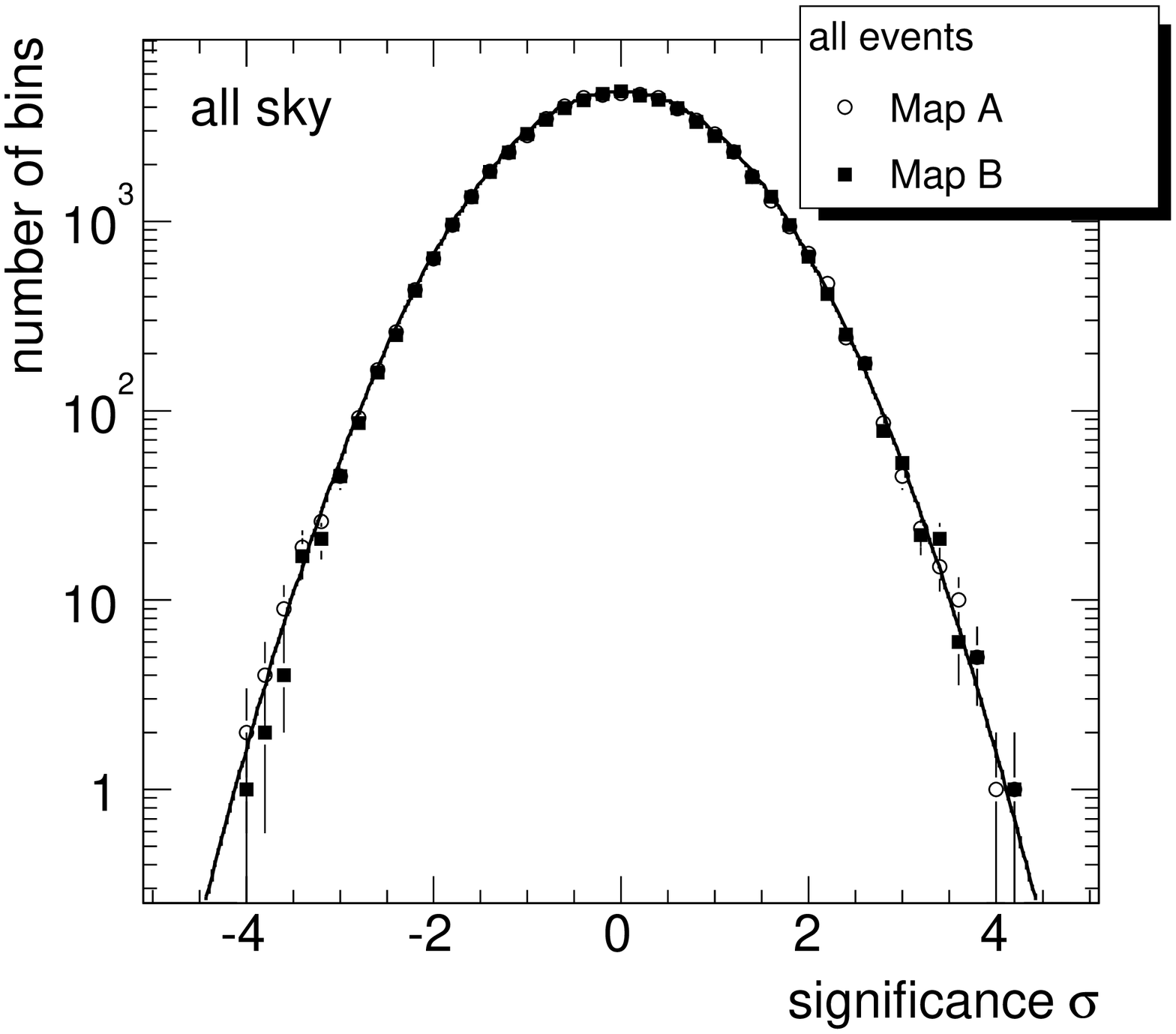}{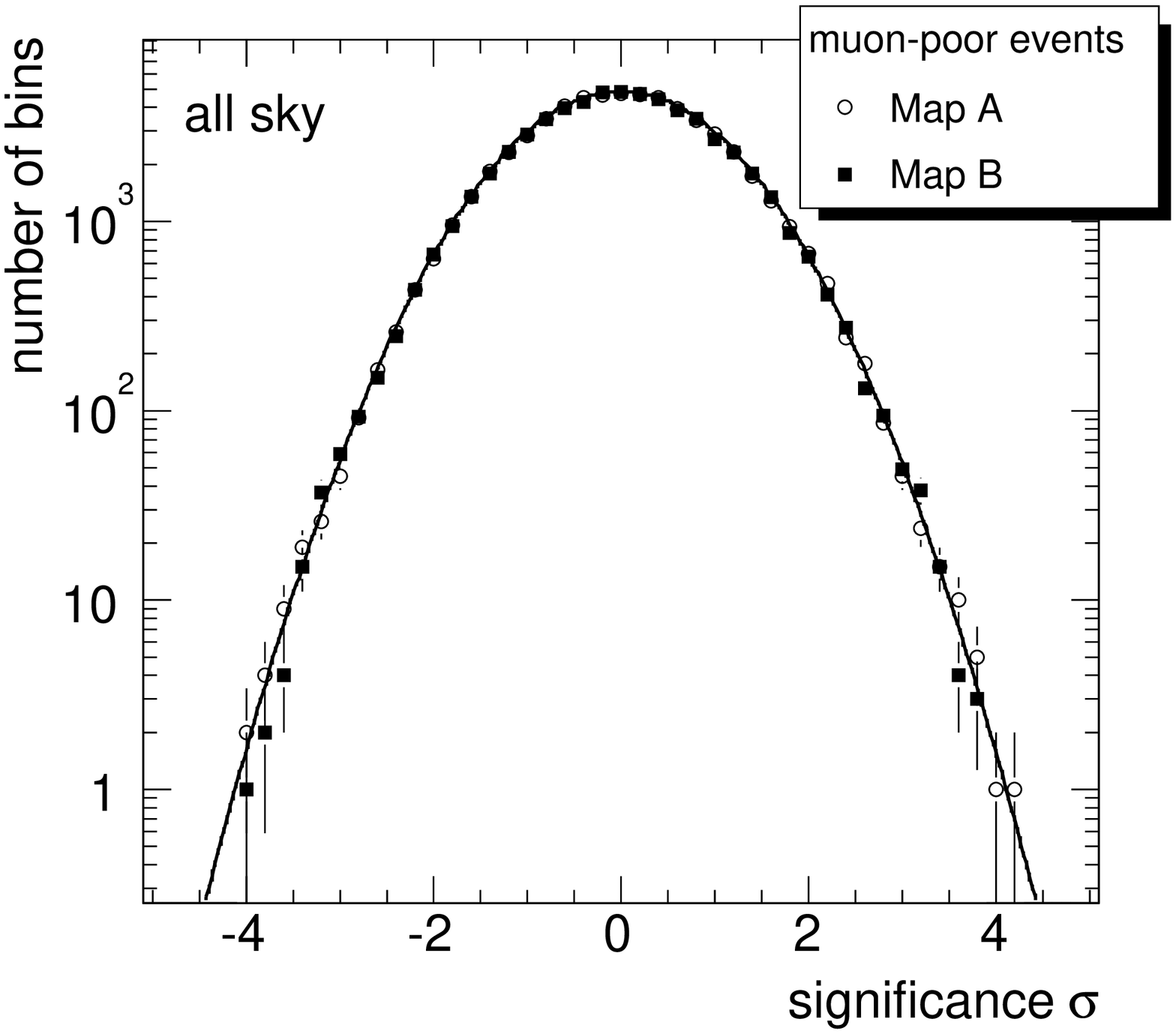}
\caption{\label{fig:suLiMa}
Distributions of the significance values from the sky maps for
data sets without (left-hand side) and with 
selection of muon-poor showers (right-hand side).
The shift of the binning grid between maps A and B is roughly half
a bin width in both directions.
The lines indicate Gaussian functions with $\mu=0$ and $\sigma=1$.}
\end{figure}

Figure \ref{fig:LiMa} shows the sky map of significance values in equatorial coordinates
for the data sample of muon-poor events.
Most of the significance values are between -1 and 1, the extreme values
are around $|\sigma|=4$.
The distributions of the significance values derived from this graph
and from maps generated from all events
are shown for two examples of binning definitions in Figure \ref{fig:suLiMa}.
The shift of the binning grid between maps A and B in Figure \ref{fig:suLiMa}
is roughly half a bin width on both axes.
Although there are deviations of about four sigma 
from the expected background 
visible in the sky map of significance values in Figure \ref{fig:LiMa},
the one-dimensional distributions show that these are expected fluctuations.
This is due to the large number of bins in the sky map ($\approx 60000$).
No unexpectedly large significance values are visible, neither in the all data nor
in the muon-poor sample.

\begin{figure}
\plotone{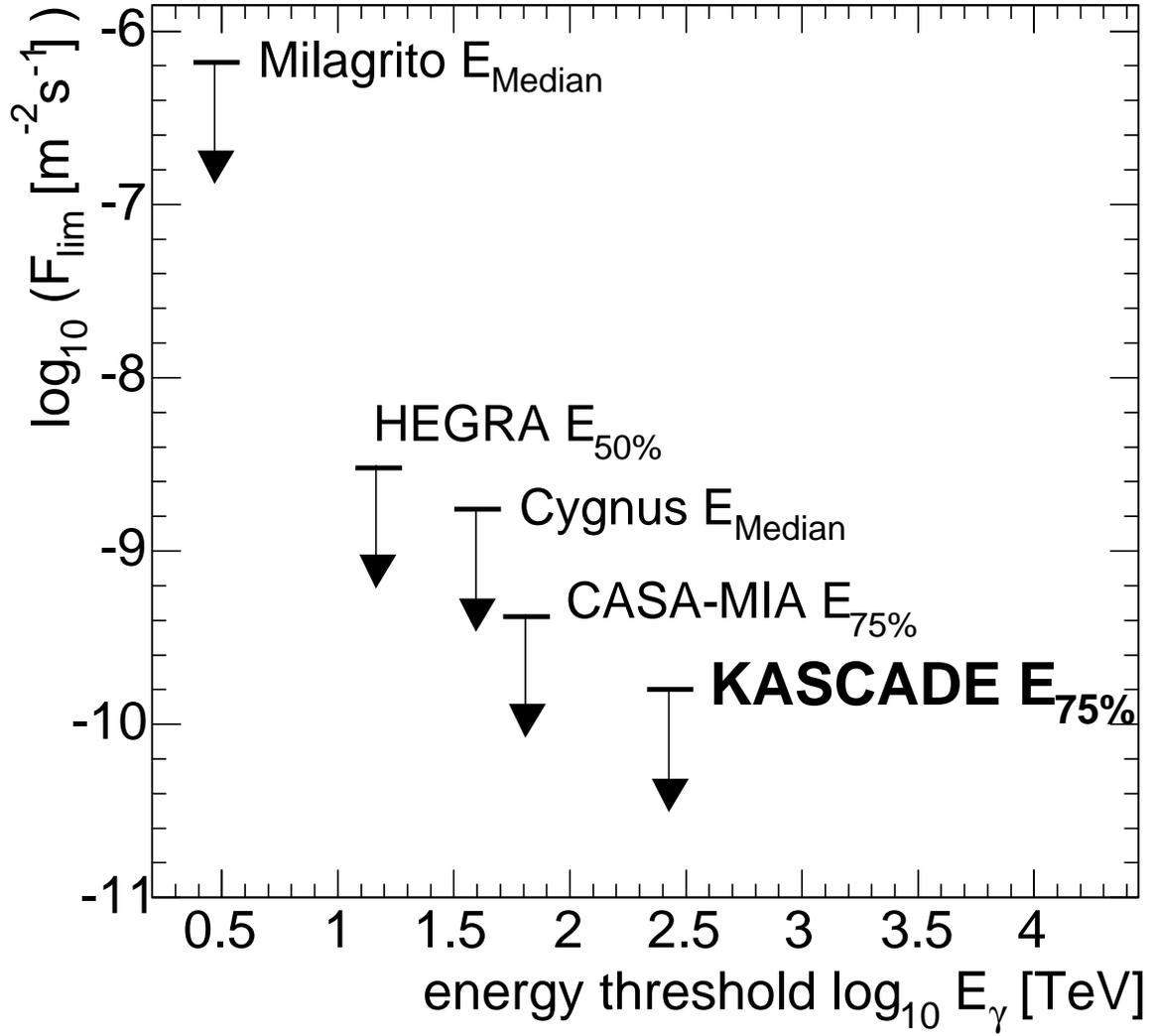}
\caption{\label{fig:flux}
90\% upper flux limit for a source moving through the zenith in
comparison with results from other experiments.
Note the different definitions of the energy threshold
\citep{Wang:2001,Aharonian:2002,Alexandreas:1991b,McKay:1993}.}
\end{figure}

The result of the upper flux limit (90\% confidence) calculated according
to equation~\ref{equ:flux}
for a source moving through the zenith is compared with results from
other experiments in Figure \ref{fig:flux}.
The different definitions of the energy threshold should be noted.
Some of the experiments use the energy, where 50\% or 75\% of the showers are detected,
others the median energy of a source moving through the zenith.
The decrease of the upper flux limits with energy reflects the
power law of the primary energy spectrum.
The typical upper flux limit determined for the present data set is
about $(0.9-5) \times 10^{-10}$~m$^{-2}$~s$^{-1}$ at an energy of about 300~TeV.
This is roughly 1-2 orders of
magnitude larger than the
flux of the Crab-Nebula extrapolated up to this energy.

\subsection{Analysis of point source candidates}
\label{sec:results2}

\begin{figure}
\plottwo{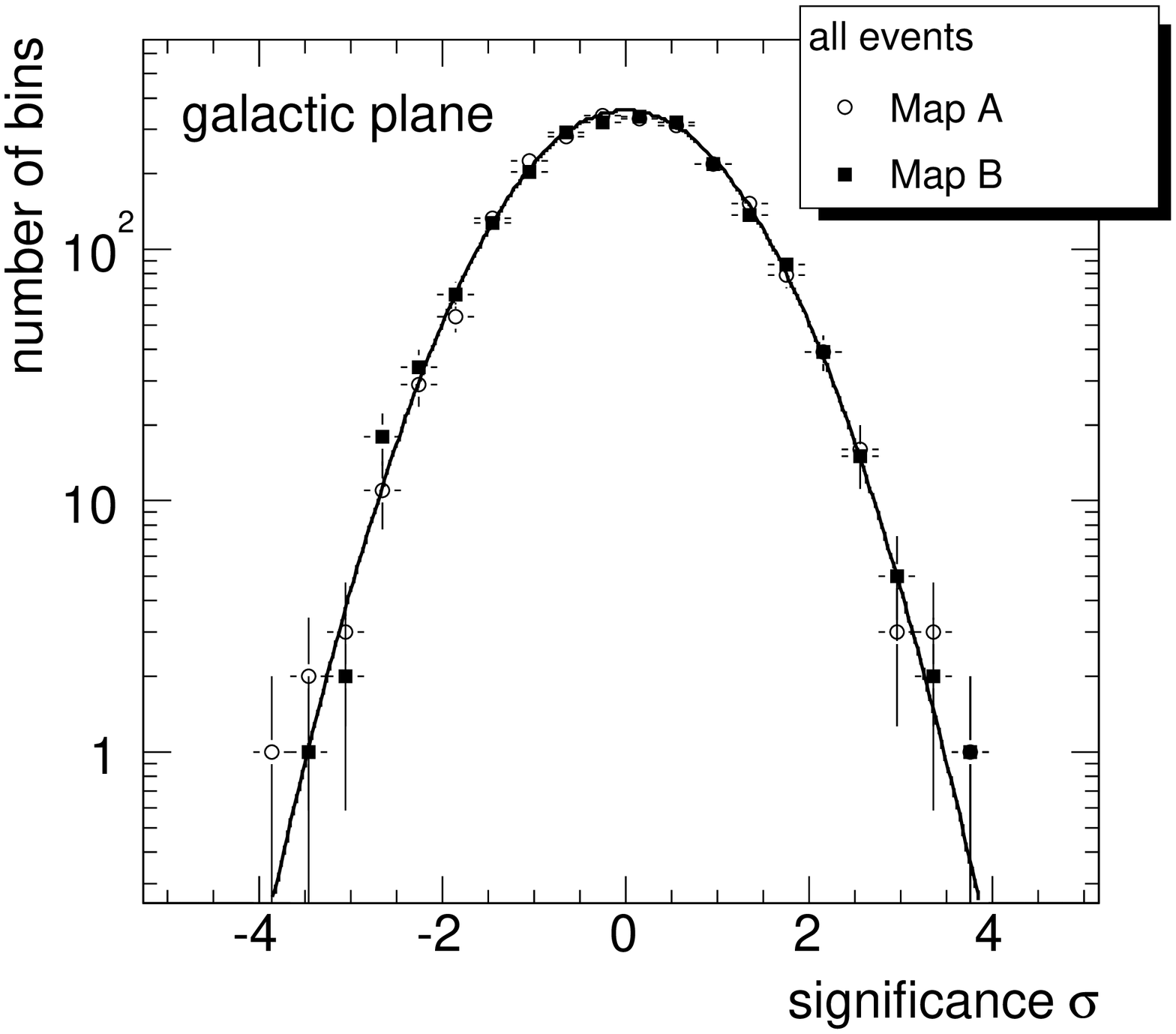}{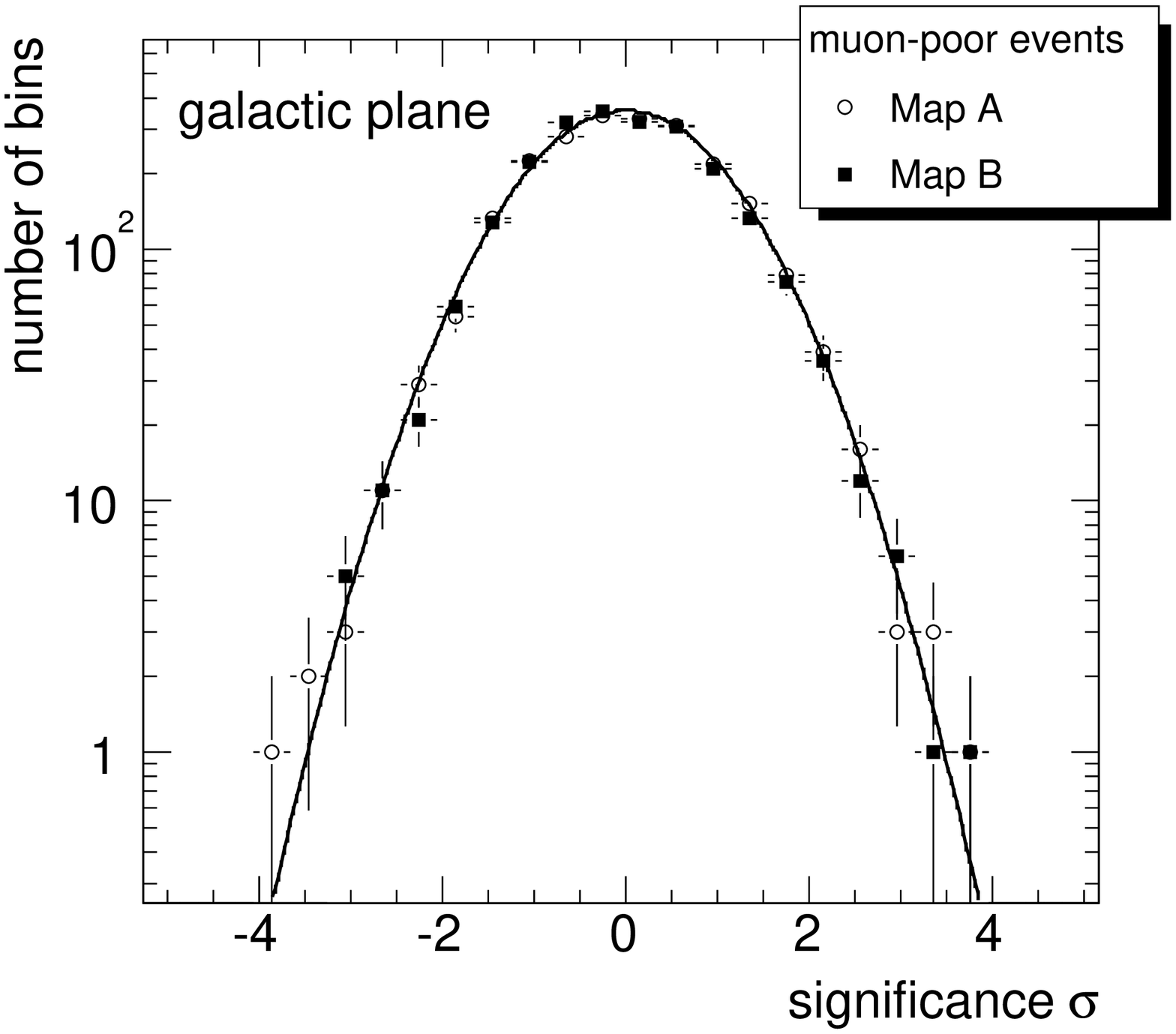}
\caption{\label{fig:suMWLiMa}
Distributions of significance values from a band of $\pm 1.5$\Grad \ around the galactic
plane for data sets without (left-hand side) and with selection 
on muon-poor showers (right-hand side).
The shift of the binning grid between maps A and B is roughly half
a bin width in both directions.
The lines indicate Gaussian functions with $\mu=0,\sigma=1$.}
\end{figure}

The most probable candidates for cosmic-ray point sources
are expected to be of galactic origin.
The distribution of the significances in a band-like region with
a width of $\pm$1.5\Grad \ around the galactic plane is shown
in Figure \ref{fig:suMWLiMa}.
Again, no large unexpected significance values have been detected,
i.e. the distributions have Gaussian shapes.

The results of the analysis of discs (radius 0.5\Grad )
centered at the position
of presently known TeV $\gamma$-ray sources
are listed in Table \ref{tab:CirQuellen}.
There is no indication for an excess from any of these source candidates, 
the largest positive significance is $1.69 \sigma$.

\cite{Chilingarian:2003} reported recently the detection of a source of
high-energy cosmic rays in the Monogem Ring,
from right ascension 7.$^{\mathrm{h}}$5, declination 14\Grad \ (750+14).
This is just outside the declination range considered here, but changing
slightly the quality cuts extends the visible sky 
of KASCADE sufficiently.
In the KASCADE data set there are in total 742 events with a maximum
distance of 0.5\Grad \ to the suggested location.
The number of expected events is 716, this corresponds to an
excess significance of 0.94 $\sigma$ and an upper flux limit
of $3\times 10^{-10}$ m$^{-2}$ s$^{-1}$.
Furthermore, no significant signal ($\sigma = 1.45$, 785 measured and 745 expected events)
is seen from the region around
PSR B0656+14 in the center of the Monogem Ring, which was proposed by \cite{Thorsett:2003} as
a nearby cosmic-ray source.


\section{Autocorrelation of showers $\mathbf{E_0 >}$ 80 PeV}
\label{sec:auto}

The most energetic extensive air showers 
measured by the \mbox{KASCADE} experiment correspond to
primary energies around 80-140~PeV.
The data set is analyzed
using an estimator of the autocorrelation function
according to \cite{Landy:1993},
which has the advantage of a Poisson-like error calculation.
It describes essentially the ratio of
the probabilities to find a pair of showers
separated by a certain angular distance $\mathrm{d}\phi$ in
the measured data set and the one derived from an
isotropic distribution:
\begin{equation}
1+w_4(\mathrm{d}\phi)=(DD-2DR-RR)/RR
\end{equation}
$DD,DR,RR$ denote the angular distance
distributions
of data-data, data-random and random-random events. 
To reproduce an isotropic background,
random directions $R$
are generated from the measured directions $D$ again
using the time-shuffling
technique 
averaging this time over 1000 artificial data samples.

\begin{figure}
\plottwo{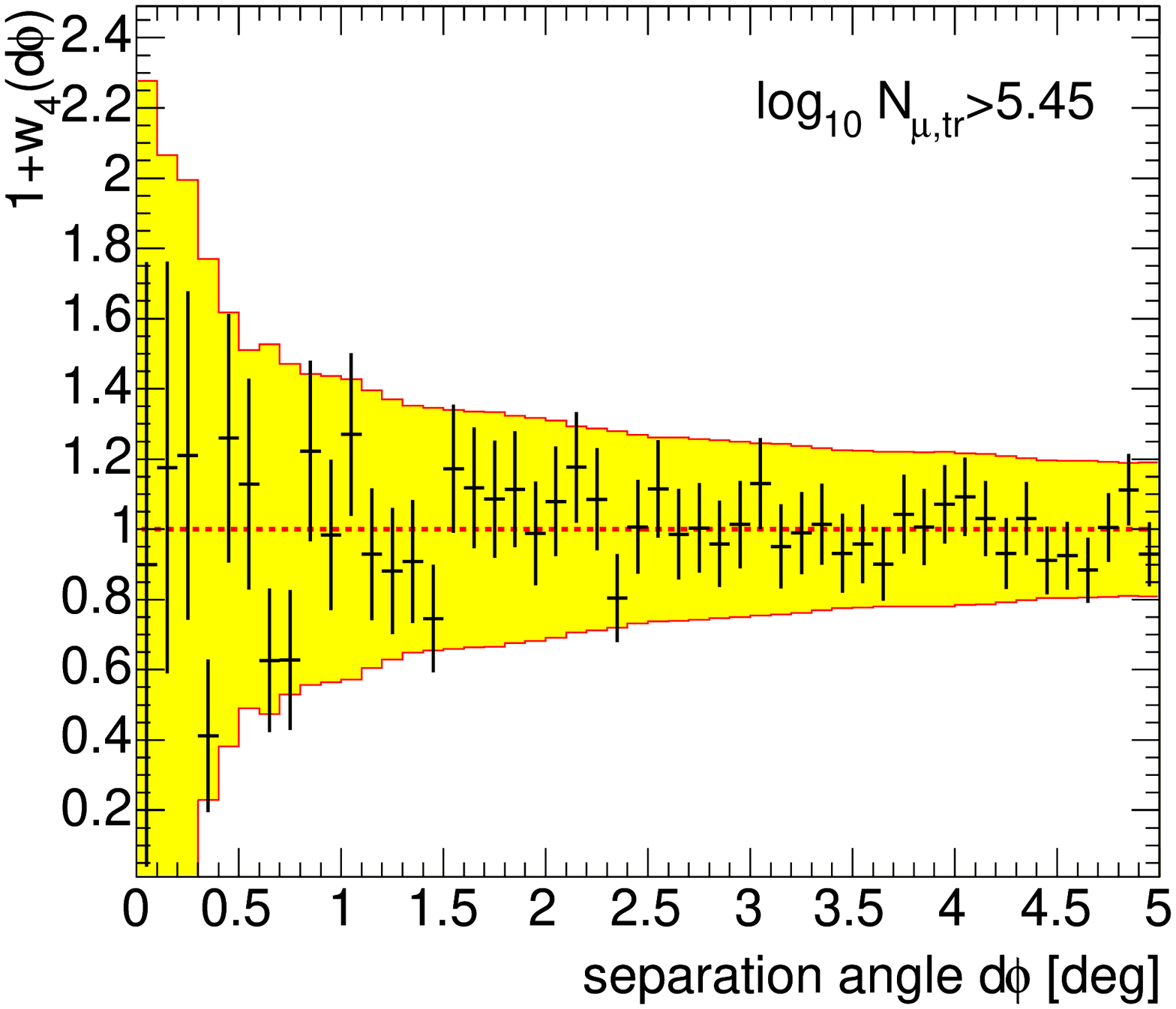}{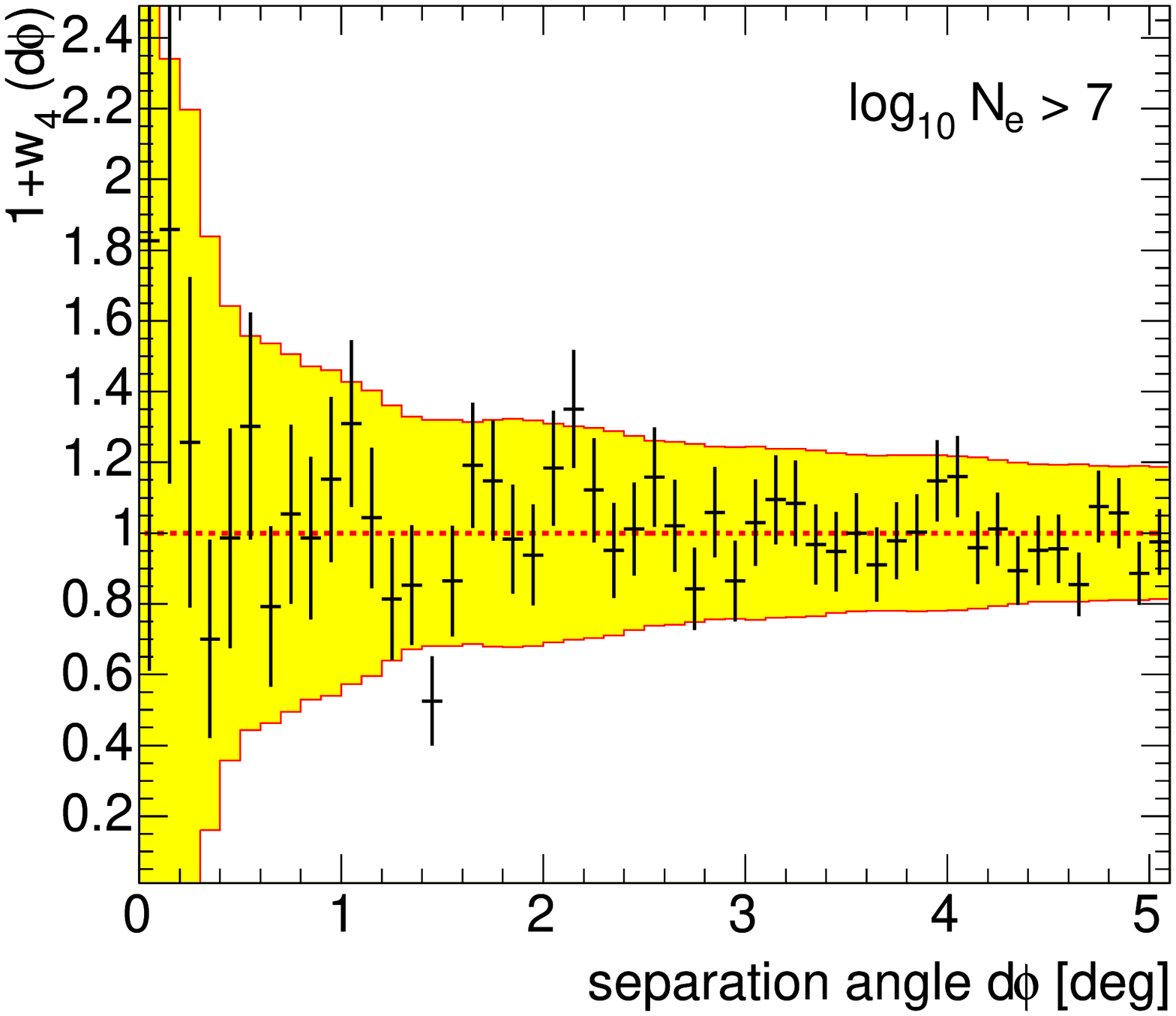}
\caption{\label{fig:AutoCorr}
Autocorrelation functions $1+w_4(\mathrm{d}\phi)$ of showers with 
$\log_{10} N_{\mu,tr}>5.45$ (left-hand side) and $\log_{10} N_e > 7$ (right-hand side).
The shaded areas indicate the two sigma (95\%) confidence regions.}
\end{figure}

Figure \ref{fig:AutoCorr} shows the $1+w_4(\phi)$ distributions
for the 1000 largest showers selected by truncated muon numbers $N_{\mu,tr}$
(left-hand side) and selected by electron numbers $N_e$ (right-hand side).
The number of muons is a better estimator
of the primary energy compared to the electron number $N_e$,
but a selection of showers by their muon number suppresses possible
$\gamma$-ray induced showers.
No significant deviation from the isotropic expectation,
which is exactly one, is detected.
A potential point source would be visible in an enhancement of
values below 0.2\Grad , which corresponds to the
angular reconstruction accuracy.
The points are inside the estimation of the
two sigma (95\%) confidence regions indicated by the shaded area.
These results are expected, since a clustering of the showers
at this primary energies is unlikely due to
the low flux and large attenuation of $\gamma$-rays or
the short decay length of neutrons of only about 1 kpc.


\section{Summary}

A search for small scale anisotropies in a data set of about 47 Mio
extensive air showers
measured with the KASCADE experiment has been presented.
The arrival direction of all events in this data set as well
as a subset of muon-poor events, i.e.~extensive air showers
which are more similar to $\gamma$-ray induced showers, 
are analysed.
No evidence for any point-like source has been found,
which is in accordance with theoretical expectations.
This is valid for the sky survey as well as for
a detailed analysis of 
a region around the galactic plane and 
known TeV $\gamma$-ray sources.
Upper limits for the detection of point-like sources are determined to be
around $10^{-10}$ m$^{-2}$ s$^{-1}$.
Also no clustering of the arrival direction for showers 
with primary energies above 80 PeV is visible.


\acknowledgments

The authors would like to thank the members of the engineering and technical staff of the KASCADE
collaboration who contributed with enthusiasm and commitment to the success of the experiment. 
The KASCADE experiment is supported by the German Federal Ministry of Education and Research
and was embedded in collaborative WTZ projects between Germany and Romania (RUM 97/014) and
Poland (POL 99/005) and Armenia (ARM 98/002). 
The Polish group acknowledges the support by KBN grant no. 5PO3B 13320.

\clearpage

%

%

%

%

%

\clearpage

\thispagestyle{empty}

\begin{deluxetable}{lrrrrrrrccrrrc}
\rotate
\tablewidth{0pt}
\tablecaption{\label{tab:CirQuellen}
Results of the analysis of discs (radius 0.5\Grad ) centered at the positions
of currently known TeV $\gamma$-ray sources.}
\tablehead{
& & & & & \multicolumn{4}{l}{all events}  & & \multicolumn{4}{l}{muon-poor events} \\
\cline{6-9} \cline{11-14} \\
source name   & RA & Dec & $T_{obs}$ & $E_{75\%}$ & $N_D$ & $N_{B}$ & $\sigma$          & $\log_{10} F_{lim}$ & & $N_D$ & $N_{B}$ & $\sigma$ & $\log_{10} F_{lim}$ \\
              &      &      & (h)       & (TeV)      &       &          &                   & (m$^{-2}$s$^{-1}$)&  &       &           &         &(m$^{-2}$s$^{-1}$) 
}
\startdata
Crab Nebula    & $05^{\mathrm{h}}34^{\mathrm{m}}31^{\mathrm{s}}$ & $+22$\Grad $01$' & 5843    & 510  & 2680         & 2695     & -0.29         & -9.50  & & 927          & 925  & 0.08  & -9.81     \\
Cas A           & $23^{\mathrm{h}}23^{\mathrm{m}}26^{\mathrm{s}}$ & $+58$\Grad $48$' & 11077   & 300 & 11069        & 11204    & -1.26         & -9.32  & & 3487         & 3483 & 0.06  & -9.96     \\
MRK 421        & $11^{\mathrm{h}}04^{\mathrm{m}}27^{\mathrm{s}}$ & $+38$\Grad $12$' & 8610    & 315 & 8497         & 8650     & -1.64         & -9.35  & & 2666         & 2667 & -0.02 & -9.64     \\
MRK 501        & $16^{\mathrm{h}}53^{\mathrm{m}}52^{\mathrm{s}}$  & $+39$\Grad $45$' & 8969    & 318  & 9289         & 9246     & 0.45          & -9.18  & & 2871         & 2855 & 0.30  & -9.62     \\
1ES2344+514    & $23^{\mathrm{h}}47^{\mathrm{m}}04^{\mathrm{s}}$ & $+51$\Grad $42$' & 10275   & 280 & 11733        & 11549    & 1.69          & -9.07  & & 3681         & 3599 & 1.35  & -9.52    \\ 
1ES1959+650    & $19^{\mathrm{h}}59^{\mathrm{m}}60^{\mathrm{s}}$ & $+65$\Grad $09$' & 11664   & 340 & 9604         & 9711     & -1.08         & -9.39  & & 2990         & 3096 & -1.90 & -9.80     \\
3C66A          & $02^{\mathrm{h}}22^{\mathrm{m}}40^{\mathrm{s}}$ & $+43$\Grad $02$' & 9295    & 288 & 10198        & 10111    & 0.85          & -9.14  & & 3139         & 3108 & 0.55  & -9.67    \\
H1426+428      & $14^{\mathrm{h}}28^{\mathrm{m}}32^{\mathrm{s}}$  & $+42$\Grad $40$' & 9322    & 285 & 10108        & 10149    & -0.41         & -9.23  & & 3133         & 3149 & -0.28 & -9.66     \\
TeV J2032+4130 & $20^{\mathrm{h}}32^{\mathrm{m}}07^{\mathrm{s}}$  & $+41$\Grad $30$' & 9110    & 285 & 9724         & 9668     & 0.56          & -9.16  & & 3004         & 2998 & 0.11  & -10.18   \\
\enddata
\tablecomments{
The meaning of the columns is (from left to right): right ascension (J2000), declination (J2000),
total observation time of this sky region ($T_{obs}$), threshold energy ($E_{75\%}$),
number of measured events in 
this source region ($N_D$), number of background events ($N_{B}$), 
significance $\sigma$
of deviation of $N_D$ from $N_{B}$, and the upper flux limit ($F_{lim}$) calculated
with Eq.~\ref{equ:flux}.
The values are given for the whole data set and the selection of muon-poor events.
}
\end{deluxetable}
\end{document}